\documentstyle[prl,aps,multicol]{revtex}
\begin{document}
\tighten

\draft
\title{Effective $\sigma$ Model Formulation for Two
Interacting Electrons in a Disordered Metal}

\author{Klaus Frahm\cite{kf}, Axel M\"uller--Groeling\cite{amg} and
        Jean-Louis Pichard}

\address{Service de Physique de l'\'Etat condens\'e,
        Commissariat \`a l'Energie Atomique Saclay,
        91191 Gif-sur-Yvette, France}

\date{\today}

\maketitle

\begin{abstract}
We derive an analytical theory for two interacting electrons in
a $d$--dimensional random potential. Our treatment is based on an
effective random matrix Hamiltonian. After mapping the problem on a
nonlinear $\sigma$ model, we exploit similarities with the theory of
disordered metals to identify a scaling parameter, investigate the level
correlation function, and study the transport properties of the system.
In agreement with recent numerical work we find that pair propagation is
subdiffusive and that the pair size grows logarithmically with time.
\end{abstract}

\begin{multicols}{2}
\narrowtext

In the theory of Anderson localization one is accustomed to
characterize localized single particle states by their typical
extension. This so--called localization length is also believed
to be the relevant length scale separating metallic and insulating
behavior. However, it has become clear recently that two interacting
electrons can form a pair that propagates up to a scale $L_2$,
far beyond the
one--particle localization length $L_1$. This effect, originally
proposed by Shepeleyansky \cite{shep1} and subsequently investigated, made
more precise and generalized in a series of papers
\cite{imry,numerics}, has added
an important aspect to the theory of localization. Inhowfar this
phenomenon is amenable to experimental verification is presently under
active consideration in the community. The theoretical information
accumulated so far is, however, mainly of either qualitative or
numerical nature. It is the purpose of this letter to develop an
analytical theory of the effect
for length scales $L > L_1$, where the
interaction--assisted pairs are well defined.

Starting from an effective
Hamiltonian we derive a nonlinear $\sigma$ model describing the
problem in arbitrary dimensions.
This enables us to address a number of interesting issues. We
identify a certain scaling parameter, namely
an effective ``pair conductivity''
$\sigma_{eff}$, which coincides in one dimension with the pair
localization length $L_2\sim \sigma_{eff}\sim L_1^2$. In the
perturbative regime, the $\sigma$ model gives rise to
subdiffusive dynamics, where both the
diffusion constant
${\cal D}_{eff}(\omega)$ of the pairs
and the local density of diffusing pair states
$\nu_{eff}(\omega)$ depend on frequency.
As a consequence,
we find that (for a certain time scale) the pair
size grows logarithmically with time and that the diffusion is
supressed by a similar logarithmic factor. These results have some
very interesting relations with recent numerical findings
\cite{shep3,shep4} for two ``interacting'' kicked rotators.
Finally, we identify two regimes with different spectral statistics
separated by a pair Thouless energy $E_c^{(2)}$.

The first step is to motivate the effective random matrix Hamiltonian
(ERMH) that serves as our starting point.
We consider two interacting electrons on a lattice in $d$
dimensions with site diagonal disorder and nearest neighbor hopping
elements. We assume that all one--electron eigenstates are localized with a
typical localization length $L_1\gg 1$ (all length scales are measured
in units of the lattice spacing).
For $d=1,2$ this condition implies
sufficiently weak disorder while for $d>2$ the
disorder strength should be slightly above the critical disorder $W_c$.
The localized one electron states
are denoted by
$\varphi_\rho(r)$, where the index
$\rho$ refers to the ``center'' of the state.
The Hamiltonian, expressed
in the basis of symmetrized (i.e. both electrons have opposite spin)
one--electron
product states $|\rho_1 \rho_2\rangle$ with $\langle
r_1r_2|\rho_1\rho_2\rangle =
[\varphi_{\rho_1}(r_1)\varphi_{\rho_2}(r_2)+
\varphi_{\rho_1}(r_2)\varphi_{\rho_2}(r_1)]/\sqrt{2}$, reads
\begin{equation}
\label{eq:1}
H_{\rho_1\rho_2,\,\rho_3\rho_4}=
(\varepsilon_{\rho_1}+\varepsilon_{\rho_2})\delta_{\rho_1,\rho_2}\,
\delta_{\rho_3,\rho_4}\,
+Q_{\rho_1 \rho_2,\, \rho_3 \rho_4} \quad.
\end{equation}
Here, the $Q_{\rho_1 \rho_2,\, \rho_3 \rho_4}$ are the interaction matrix
elements expressed in the basis of the
product states $|\rho_1 \rho_2>$ and
$\varepsilon_\rho$ is the one--electron energy of the localized state
$\varphi_\rho(r)$.
We emphasize that we consider the dynamics of the pairs in the basis
of the above product states. Only for scales $L>L_1$ can this be
easily interpreted as motion in real space.
For a local Hubbard interaction with interaction
strength $U$ we have \cite{shep1}:
\begin{equation}
\label{eq:2}
Q_{\rho_1 \rho_2,\, \rho_3 \rho_4}
=2U\sum_r \varphi_{\rho_1}(r)\,\varphi_{\rho_2}(r)\,\varphi_{\rho_3}(r)\,
\varphi_{\rho_4}(r)\quad.
\end{equation}
These interaction matrix elements become exponentially small
whenever any two of
the ``positions'' $\rho_j$
differ by much more than $L_1$.
Shepelyansky mapped \cite{shep1}
(for $d=1$) the Hamiltonian (\ref{eq:1})
on a random band matrix with a superimposed,
strongly fluctuating diagonal matrix. To do so he
employed two crucial approximations:  First he neglected all badly
coupled pair states (those with $|\rho_1-\rho_2|\ge L_1$) and
second he assumed the remaining coupling elements (\ref{eq:2}) to be
independent,
equally distributed random variables.
With the
ansatz $\varphi_\rho(r)\sim e^{-|r-\rho|/ L_1}\,a_\rho(r)
/L_1^{d/2}$, where $a_\rho(r)$ is a random variable with
$\langle a_\rho(r)\rangle =0$ and $\langle a_\rho(r)
a_{\tilde\rho}(\tilde r)\rangle =\delta_{\rho \tilde\rho}\,
\delta_{r \tilde r}$,
and using the central limit theorem one gets the
estimate \cite{shep1,imry}
$\left\langle Q_{\rho_1 \rho_2,\, \rho_3 \rho_4}^2\right\rangle \sim
U^2 L_1^{-3d}$ for the well--coupled states.
Our ERMH defined below only relies on the second of the above
assumptions: {\it All} pair states are taken into account, but the
coupling matrix elements (\ref{eq:2}) will still be taken to be independent
Gaussian variables. This latter point is indeed quite a serious
simplification since the correlations between different coupling
matrix elements are neglected.
We believe, however, that essential
physics can be learnt even without paying attention to this
refinement.
In particular, the strongly fluctuating diagonal elements tend to
eliminate the effect of those correlations.

Let us consider as configuration space a $2d$--dimensional lattice with
two coordinate vectors
$R\equiv \rho_1+\rho_2$ and $j\equiv(\rho_1-\rho_2)/2$
corresponding to twice the center of mass and half the distance of the two
electrons, respectively.
Our ERMH,
${\cal H}=\hat\eta+\hat \zeta$, consists of a strongly fluctuating diagonal
part $\hat\eta$ with entries $\eta_R^{\,j}$ and of an interaction induced
coupling matrix $\hat\zeta$. The $\eta_R^{\,j}$
correspond to the one--electron energies $\varepsilon_{\rho_1}+
\varepsilon_{\rho_2}$ in
(\ref{eq:1}). We take them to be independent random variables with the
distribution function
$\rho_0(\eta)$. We typically have
$\rho_0(\eta)\simeq 1/(2W_b)$ for $|\eta|\le W_b$, where $W_b$
is the bandwidth of the disordered one--electron Hamiltonian.
The matrix elements of $\zeta$ are independent Gaussian random
variables with zero mean and variance
\begin{displaymath}
\Bigl \langle \left(\hat\zeta_{R \tilde R}^{j \tilde j}\right)^2\Bigr
\rangle =\frac{1}{2}(1+\delta_{R \tilde R}\,\delta_{j \tilde j})
\,a(|R-\tilde R|)\, v(j)\,v(\tilde j)\quad.
\end{displaymath}
Here $a(|R-\tilde R|)$ and $v(j)$ are
smooth functions decaying exponentially
on the scale $L_1$. We need not specify their particular form,
it is sufficient to know their typical behavior
\begin{eqnarray}
\label{eq:4}
a(|R|) & \sim &
\left\{\begin{array}{ll}
U^2 L_1^{-3d} &\ ,|R|\lesssim L_1\ , \phantom{\Big|}\\
U^2 L_1^{-3d} e^{-2|R|/L_1} &\ ,
|R|\gg L_1\ , \phantom{\Big|}\\
\end{array}\right.\\
\label{eq:5}
v(j)& \sim &
\left\{\begin{array}{ll}
1 &\ ,|j|\lesssim L_1\ , \phantom{\Big|}\\
e^{-4|j|/L_1} &\ ,|j|\gg L_1\ , \phantom{\Big|}\\
\end{array}\right.
\end{eqnarray}
which is justified by (\ref{eq:2}) and the exponential decrease (on the
scale $L_1$) of the localized eigenfunctions.
The function $a(|R-\tilde R|)$ describes how the coupling strength decreases
exponentially if the distance of the centers of masse increases,
wheras $v(j)$ describes how the increasing size of the pair states reduces the
coupling.

To investigate
the spectral statistics, the transport and the
localization properties of the ERMH we apply the supersymmetric
technique \cite{efetov,vwz}. In the following, we choose a fixed
realization of the diagonal elements $\eta_R^j$ and restrict the ensemble
average to the random variables in $\hat\zeta$. This particular
average is denoted by
$\langle \cdots\rangle_\zeta$. We consider the generating functional
\begin{displaymath}
F(J) =
\Bigl\langle \int {\textstyle
D\psi\,\exp\left[\frac{i}{2}\bar\psi
(E-{\cal H} +(\frac{\omega}{2}+i\varepsilon)\Lambda+J)\psi\right]
}\Bigr\rangle_\zeta \ .
\end{displaymath}
Here,  $\psi$ is a supervector with components $\psi_j(R)$, which are
themselves
8--dimensional supervectors with entries $z_1,\bar z_1,\chi_1,\bar\chi_1,
z_2,\bar z_2,\chi_2,\bar\chi_2$, where the $z_\nu$ ($\chi_\nu$) are complex
bosonic (fermionic) variables.
The diagonal matrix $\Lambda$ has an equal number of eigenvalues
$+1$ and $-1$ and describes
the grading into advanced and retarded Greens' functions.
Furthermore, $\omega$ is a
frequency, $J$ a source matrix, and $\bar\psi$ is given by $\bar\psi
=\psi^\dagger\Lambda$.

Choosing $J$ appropriately
\cite{vwz,iwz} and taking derivatives of $F(J)$
with respect to $J$ at $J=0$,
one obtains ensemble averages of arbitrary products of Green's
functions, in terms of which the above mentioned properties can be
studied.
Skipping most of the technical detail (for more information see
\cite{efetov,vwz}), we derive a nonlinear $\sigma$ model from the functional
$F(J)$. The physical implications are then discussed by comparing
our formulation of the present problem with the supersymmetric
description of a disordered metal given by Efetov \cite{efetov}.
Performing the ensemble average, we get
\begin{eqnarray}
\nonumber
\langle{\textstyle \exp(-\frac{i}{2}\bar\psi \zeta\psi)}\rangle_\zeta&=&
\exp\biggl(-\frac{1}{8}\sum_{R,\tilde R}a(|R-\tilde R|)\\
\label{eq:7}
&&\times\mbox{str} [K(R)\, K(\tilde R)]\biggr) \quad,
\end{eqnarray}
where $K(R)$ is a $8\times 8$ super matrix  given by
$K(R)=\sum_j v(j)\,\psi_j(R)\,\bar\psi_j(R)$.
The quartic term
(\ref{eq:7}) can be decoupled in the usual way \cite{efetov,vwz} by a
Hubbard-Stratonovich transformation. This introduces a functional integration
over a
field of $8\times 8$ super matrices $\sigma(R)$ with the same symmetries as
$K(R)$. Proceeding \cite{framg2} in analogy to \cite{framg1,fyodorov2},
we apply a saddle point approximation, which is justified in the limit
$L_1\gg 1$. Then we put $\sigma(R)=\Gamma_0
+i\Gamma_1 Q(R)$, where $Q(R)$ is an element of the orthogonal
$\sigma$ model space \cite{efetov,vwz} and fulfills the nonlinear constraint
$Q(R)^2=1$.
The quantities $\Gamma_0$ and $\Gamma_1$
are determined by the implicit equation
\begin{eqnarray}
\nonumber
\Gamma_0+i\Gamma_1\Lambda & = & -B_0\sum_j v(j) \int d\eta\ \rho_0(\eta)\\
\nonumber&&\times
[E-\eta+i\varepsilon\Lambda+\frac{1}{2}v(j)
(\Gamma_0+i\Gamma_1\Lambda)]^{-1}\quad,
\end{eqnarray}
having the approximate solutions (in the limit $L_1\gg 1$):
\begin{eqnarray}
\nonumber
\Gamma_0 & \simeq & -B_0 S\,{\cal P}\int d\eta\ \rho_0(\eta)
\frac{1}{E-\eta}\quad,\\
\nonumber
\Gamma_1 & \simeq & \pi B_0 S\,\rho_0(E)\quad.
\end{eqnarray}
Here, $B_0=\sum_R a(|R|)$, $S=\sum_j v(j)$,
and ${\cal P}\int d\eta\ (\cdots)$ denotes a principal value
integral. From this and (\ref{eq:4},\ref{eq:5}) we find the estimate
$\Gamma_1\sim U^2/(W_b\, L_1^{d})\ll W_b$
if $U$ and $W_b$ are of the same order of magnitude.
The matrix $A_{R,\tilde R}=a(|R-\tilde R|)$ defines a
$d$--dimensional generalization of a random band matrix \cite{fyodorov}
with the typical bandwidth $L_1$.
As a consequence of the Hubbard--Stratonovich transformation the
coupling in the bilinear term in the $Q$--field is given by the
inverse,
$(A^{-1})_{R,\tilde R}$. Therefore mainly the slow modes
\cite{efetov,fyodorov} (i.e. small momenta) contribute to the
functional integral. Performing a standard gradient expansion
\cite{efetov,fyodorov} and going
over to the continuum limit, we can express the
generating functional as $F(J)=\int DQ\,
\exp[-{\cal L}_2[Q]-\Delta {\cal L}(J)]$. The effective action is
given by
\begin{eqnarray}
{\cal L}_2[Q] & = & \int dR\ \mbox{str}\left[
-\frac{\Gamma_1^2 B_2}{8 B_0^2}[\nabla_R\, Q(R)]^2
+f_R(Q(R))\right]
\nonumber\\
f_R(Q)&=&\frac{1}{2}\sum_j \mbox{str}\,\ln \Bigl( {\textstyle
E-\eta_R^j +(\frac{\omega}{2}
+i\varepsilon)\Lambda}
\nonumber\\
&&{\textstyle+\frac{1}{2}v(j)\,(\Gamma_0+i\Gamma_1 Q)}\Bigr) \quad,
\label{eq:11}
\end{eqnarray}
where $B_2=1/(2d)\sum_R R^2\,a(|R|)$ and $\Delta {\cal L}(J)$ is the
part of the action that accounts for the source matrix \cite{framg2}.
The $Q$--dependent ``potential'' $f_R(Q)$ can be written in a more convenient
form \cite{framg2}
\begin{equation}
\label{eq:12}
f_R(Q)\simeq -i(\pi/4)\,\omega\ h(\Gamma_1/\omega)
\,\rho_0(E)\ \mbox{str}(Q\Lambda)
\end{equation}
with $h(y)=iy\sum_{|j|\lesssim L_c} v(j)/[1+iy\, v(j)]$, where $L_c\approx
L_1\ln L_1$ is a cutoff length to be explained below.

Before discussing this result, we mention that
it is also straightforward to derive
the $\zeta$--averaged local density of states from the functional
$F(J)$. It has the Breit--Wigner form
\begin{displaymath}
\langle \rho_R^j(E)\rangle_\zeta =
\frac{1}{\pi}\ \frac{\frac{1}{2}\Gamma_1\, v(j)}
{[E-\eta_R^j+\frac{1}{2}\Gamma_0\, v(j)]^2 + [\frac{1}{2} \Gamma_1\,
v(j)]^2}
\end{displaymath}
with an energy width $\Gamma_1\,v(j)$ {\it that depends
on the relative coordinate} $j$.
The levels $\eta_R^j$ of the product states $|\rho_1 \rho_2>$
acquire
a finite width (or inverse life time) $\Gamma_1\,v(j)$
due to the interaction $\hat{\zeta}$.
This width is of the order of $\Gamma_1$ for
$|j|\lesssim L_1$ and decreases exponentially like $\Gamma_1\,e^{-4|j|/L_1}$
for $|j|\gg L_1$. This result
demonstrates that product-states with
($|j|\gg L_1$) have an
exponentially large life time because both electrons are
simply localized far away from each other without feeling the
Hubbard interaction.
Let us determine a critical length
$L_c$ such that inside a volume of this diameter all pair states are well
coupled.
Equating the effective level spacing
$\Delta_{eff}=W_b\,L_c^{-2d}$
with the smallest possible level width
$\Gamma_1\,e^{-2L_c/L_1}$, yields the estimate
$L_c\sim L_1\ln L_1$.
Product states with $|j|>L_c$ contribute to a discrete
point spectrum while product states with $|j|\lesssim L_c$ are
well coupled and correspond to interaction-assisted pairs of the
size $L_c$.

The dynamics of these well-coupled product-states
is conveniently described in terms of the
$\sigma$ model (\ref{eq:11}).
At this point we reiterate that our model describes diffusion
(and localization) in the space of product states
$|\rho_1\rho_2\rangle$. The translation to coordinate space is
straightforward provided we consider scales $L>L_1$, where the
$|\rho_1\rho_2\rangle$ are associated with well defined positions.
The interpretation of our $\sigma$ model is greatly facilitated
by the close similarities
between (\ref{eq:11}) and the $\sigma$ model for a {\it disordered metal}
as derived by Efetov \cite{efetov}.
Comparing (\ref{eq:11}) and (\ref{eq:12}) with
the standard $\sigma$ model in \cite{efetov}, one can
formally identify an effective
diffusion constant ${\cal D}_{eff}$
and an effective local density of (diffusing) states
$\nu_{eff}$,
both of which depend on the frequency $\omega$:
\begin{eqnarray}
\label{eq:14}
\sigma_{eff}=\nu_{eff}(\omega)\,{\cal D}_{eff}(\omega) & = &
\frac{1}{\pi}\,\frac{\Gamma_1^2\,B_2}{B_0^2}
\sim \frac{U^2}{W_b^2} L_1^2  \quad, \\
\label{eq:15}
\nu_{eff}(\omega) & = & \rho_0(E)\,h(\Gamma_1/\omega)\quad.
\end{eqnarray}
We have introduced, via the Einstein relation, a formal ``pair
conductivity''
$\sigma_{eff}$, which does {\it not} depend on frequency.
The function $h(\Gamma_1/\omega)$ can be interpreted as the number of
states contributing to the {\em diffusion}.
A detailed analysis
\cite{framg2} yields the following limiting cases:
\begin{equation}
\label{eq:16}
h(\Gamma_1/\omega)\simeq
\left\{\begin{array}{ll}
i(\Gamma_1/\omega)\,S &,\quad \Gamma_1\ll|\omega|\ ,\phantom{\Big|}\\
\left[(L_1/4)\,\ln(\Gamma_1/|\omega|)\right]^d &
,\quad \tau_c^{-1}\lesssim |\omega|\ll \Gamma_1\ ,\phantom{\Big|}\\
\left[(L_1/4)\,\ln(\Gamma_1\,\tau_c)\right]^d &
,\quad |\omega|\ll \tau_c^{-1}\ .\phantom{\Big|}\\
\end{array}\right.
\end{equation}
Here, $\tau_c^{-1}$ is the effective level spacing $\Delta_{eff}
=\Gamma_1\,e^{-2L_c/L_1}$ inside a blocks of size $L_c$.
The physical picture of diffusing pairs
makes sense in the regime $|\omega|\ll \Gamma_1$ only.
Furthermore, in formal analogy to
the condition $|\omega|\ll 1/\tau$ in a disordered metal
\cite{efetov} (where $\tau$ is the elastic scattering time),
only those product-states with $|\omega|\lesssim
\Gamma_1\,v(j)$  contribute to the diffusion. These remarks provide a
physical interpretation of (\ref{eq:15})
and (\ref{eq:16}).

The analogy between the present problem and the problem of independent
electrons in a disordered metal enables us to draw at least three
important conclusions.

First, the coupling constant
$\frac{\pi}{8}\nu_{eff}\,{\cal D}_{eff}=
\frac{\pi}{8}\sigma_{eff}$ can be identified as a universal scaling
parameter.
The corresponding scaling function is precisely the same as
that of a disordered metal provided the latter is described by the
``standard'' $\sigma$ model \cite{efetov}. In particular, the perturbative
evaluation of the $\beta$--function in $2+\varepsilon$ dimensions
\cite{efetov,wegner,hikami} is equally valid for our present problem of
{\em diffusing} or {\em localized} electron pairs (the term (\ref{eq:12})
is not affected under the
renormalization \cite{efetov}).
This first conclusion also provides a rather rigorous justification for
Imry's \cite{imry} application of the Thouless
scaling block picture \cite{thouless}.
For $d=1$  we immediately recover Shepelyansky's original result
\cite{shep1} for the pair localization length $L_2$, $L_2\sim
\sigma_{eff}\sim (U^2/W_b^2) L_1^2$.
It is important to note that this result has been obtained by
taking into account {\em all},
also the badly coupled ($|j|\gg L_1$), pair states.
In their study \cite{shep3} Borgonovi and Shepelyansky
argue that the badly coupled states should lead to a logarithmic
correction
$L_2\sim L_1^2/\ln L_1$. We cannot confirm this result.

Second, we find that the pair dynamics is subdiffusive in agreement
with recent numerical results of Borgonovi and Shepelyansky
\cite{shep3,shep4}
who study two interacting kicked rotators. In the ``pair--metallic''
regime ($ L_1 < L\lesssim L_2$ for $d=1,2$) the $\sigma$
model can be treated perturbatively as in \cite{efetov}. The relevant
diffusion propagator ${\cal R}(q,\omega)=[\sigma_{eff}\, q^2
- -i\omega\nu_{eff}(\omega)]^{-1}=[{\cal D}_{eff}(\omega)\,
\nu_{eff}(\omega)\, q^2 -i\omega\nu_{eff}(\omega)]^{-1}$ contains both
the frequency dependent diffusion constant (\ref{eq:14}) and density of
states (\ref{eq:15}). This $\omega$ dependence gives rise to subtle
modifications of standard diffusion. Instead of trying to calculate
the diffusion propagator
$\tilde{\cal R}(R,t)=(2\pi)^{-(d+1)}\int d\omega\int dq\ {\cal R}(q,\omega)
\,e^{i(qR-\omega t)}$ as a function of $t$ and $R$ we replace in a
qualitative approximation $\omega$ by $1/t$. Therefore we expect the
pairs to diffuse
according to (see also (\ref{eq:16}))
\begin{displaymath}
\langle R^2(t)\rangle \sim {\cal D}_{eff}(\frac{1}{t})\,t\sim
\left\{\begin{array}{ll}
L_1^2 &,\ t\ll\Gamma_1^{-1}\phantom{\Big|}\\
D_0\,\ln(\Gamma_1 t)^{-d}\,t & ,\ \Gamma_1^{-1}\ll t
\lesssim \tau_c \phantom{\Big|}\\
D_0\,\ln(L_1)^{-d}\,t &,\ \tau_c\lesssim t\ .\phantom{\Big|}\\
\end{array}\right.
\end{displaymath}
We have used the notation $D_0=(U^2/W_b)\,L_1^{2-d}$.
Obviously, $\langle R^2(t)\rangle$ increases weaker than linearly with
time (subdiffusion). The number of diffusing states given by the
function $h(\Gamma_1/\omega)$ in (\ref{eq:16}) also depends on time.
Putting $h(\Gamma_1/\omega) = [L_{eff}(\omega)]^d$ with
$L_{eff}(\omega)$ the effective pair size, we get
$L_{eff}(1/t)\sim L_1\,\ln(\Gamma_1 t)$ for
$\Gamma_1^{-1}\ll t\lesssim \tau_c$ and
$L_{eff}(1/t)\sim L_1 \ln L_1$ for
$\tau_c\lesssim t$.
This means that the pair size grows logarithmically with time as has
also been found numerically in \cite{shep3,shep4}.

Third, calculating the two point correlation function $Y_2(\omega)$,
we can study the level correlations of the well--coupled product-states.
The nearly localized pair states are mainly uncorrelated with the
diffusing states and among themselves. Therefore their contribution to
$Y_2(\omega)$ essentially cancels out. For a finite system
of size $L$ the diffusive
dynamics determines another energy scale,
the ``pair Thouless energy''  $E_c^{(2)}$.
In our case $E_c^{(2)}$ is given by the implicit equation
${\cal D}_{eff}(E_c^{(2)})/L^2=E_c^{(2)}$. For
$\omega < E_c^{(2)}$ the second term (\ref{eq:12})
of the action ${\cal L}_2[Q]$ dominates the level correlations.
The function $Y_2(\omega)$ can be calculated in complete analogy
with \cite{efetov} and we recover the random matrix result:
\begin{displaymath}
Y_2(\omega)=Y_2^{(GOE)}\bigl(\omega/\Delta(\omega)\bigr)
\quad,\quad |\omega|\ll E_c^{(2)} \quad.
\end{displaymath}
Here, $\Delta(\omega)=[L^d\,\nu_{eff}(\omega)]^{-1}$ is the
frequency dependent effective level spacing and $Y_2^{(GOE)}(r)$ is the
universal spectral correlation function of the Gaussian orthogonal ensemble.
For higher frequencies, i.e. $E_c^{(2)}
\ll|\omega|\ll \Gamma_1$, also the first (kinetic) term of ${\cal L}_2[Q]$ has
to be taken into account.
The necessary perturbative evaluation of $Y_2(\omega)$ proceeds
analogously to the corresponding diagrammatical calculation of $Y_2(\omega)$
given by Altshuler and Shklovskii \cite{alt1}. The result is
\begin{displaymath}
Y_2(\omega)\sim \frac{\Delta^2(\omega)}{\omega^2}\,\left(\frac{\omega}
{{\cal D}_{eff}(\omega)/L^2}\right)^{d/2}
,\ E_c^{(2)}\ll|\omega|\ll\Gamma_1 \ ,
\end{displaymath}
where ${\cal D}_{eff}(\omega)/L^2$ can be interpreted as a frequency dependent
Thouless energy setting the scale of the level correlation
function.

In conclusion, starting from an effective
Hamiltonian, we have derived a nonlinear
$\sigma$ model for two interacting
electrons in a random potential in arbitrary dimension. Exploiting the
analogy with Efetov's description of noninteracting electrons in a
disordered metal, we identified a scaling parameter and investigated the
level correlation function for the well coupled pairs. Furthermore, we
analytically confirmed the numerical result that pair propagation is
subdiffusive and that the pair size grows logarithmically with time.

Acknowledgments. We are grateful to D. Weinmann,
Y. Imry and D.L. Shepelyansky for
fruitful discussions. This work was supported by the European HCM
program (KF) and a NATO fellowship through the DAAD (AMG).

\end{multicols}

\end{document}